# Set Theory and Many Worlds

Paul Tappenden paulpagetappenden@gmail.com

7th July 2023

The 2022 Tel Aviv conference on the Many Worlds interpretation of quantum mechanics highlighted many differences between theorists. A very significant dichotomy is between Everettian *fission* (splitting) and Saunders-Wallace-Wilson *divergence*. For fission, an observer may have multiple futures, whereas for divergence they always have a single future. Divergence was explicitly introduced to resolve the problem of pre-measurement uncertainty for Everettian theory, which is universally believed to be absent for fission. Here, I maintain that there is indeed uncertainty about future observations prior to fission, so long as objective probability is a property of Everettian branches. This is made possible if the universe is a set and branches are subsets with probability measure. A universe that is a set of universes which are macroscopically isomorphic and span all possible configurations of microscopic local beäbles fulfils that role. If objective probability is a property of branches, a successful Deutsch-Wallace decision-theoretic argument would justify the Principal Principle and be part of probability theory rather than being specific to Many Worlds. Any macroscopic object in our environment becomes a set of isomorphs with different microscopic configurations, each in an *elemental* universe (elemental in the set-theoretic sense). This is similar to Many Interacting Worlds theory but the observer inhabits the set of worlds, not an individual world. An observer has many elemental bodies.

## 1. Many Faces of Many Worlds

If a Many Worlds interpretation of quantum mechanics is ever to become generally accepted there first has to be agreement on what *the* Many Worlds interpretation is, which is very far from being the case. There's even dispute about what to call it; are we to think in terms of a single branching world or a partitioning multiplicity of worlds? Some theorists work with the Heisenberg picture and a basic ontology of operators, others work with the Schrödinger picture and a basic ontology of wavefunctions. On both approaches there's scope for arguing that microscopic local beäbles are needed for a satisfactory physical ontology.

Within the diversity of views there's a fundamental dichotomy which I aim to resolve here. It's between the ideas that an observer may have a multiplicity of futures or always has a single future. Everett wrote of *splitting* in quantum measurement situations and it has generally been accepted that a well-informed observer cannot be uncertain about their future observations prior to Everettian fission. In an attempt to introduce pre-measurement uncertainty, Simon Saunders and David Wallace developed versions of Many Worlds theory that reject the concept of splitting, which was arguably Everett's key idea. There shall be more on this in the following section.

To begin with, I will address the thorny matter of understanding the relationship between probability and uncertainty. This will lead to the conclusion that pre-measurement uncertainty exists for a fission interpretation of branching, where an Everettian observer splits into observers seeing different outcomes. The only reason why that feels counterintuitive is that we've inherited a folk metaphysics which interprets future probabilities as properties of alternative possibilities. It's that which stands in the way of interpreting probabilities as properties of future coexistent actualities. A thought experiment helps to sugar this pill.

Understanding uncertainty as a cognitive state of assigning partial degrees of belief to coexistent futures requires assigning objective probabilities to those futures equal to the absolute squares of their quantum amplitudes. What's needed is an account of how this *branch weight* can be understood to *constitute* objective probability. I shall argue that it can do so if understood to be a subset measure. This leads to interpreting the universal wavefunction as being a set of deterministic universes which contain microscopic local beäbles. Objects in our environment become sets of objects which are macroscopically isomorphic but differ in their microscopic configurations. They are set-theoretically extended in *configuration space*, so to speak.

The result is a set-theoretic metaphysics for quantum mechanics which incorporates Everettian fission and microscopic local beäbles. It opens the way to new physics if the interaction between the universes which are the set-theoretic elements or our universe is the source of phase relations. After discussing spin, separability and locality in the context of this metaphysics, I close with further reflections on Lev Vaidman's *World Splitter* and its implications.

## 2. Probability and Uncertainty

For classical mechanics, all physical processes are regarded as deterministic. The idea of there being a mind-independent, i.e. *objective*, probability can only be applied to the determination of initial conditions, relegated to an inscrutable past. Probability arises, as in statistical mechanics, from the epistemic condition of ignorance on the part of observers. Lack of Laplacian omniscience as to the exact positions and momenta of particles entails that perfect prediction is impossible, so epistemic probabilities are assigned to fictional *possibilities* on the basis of statistical evidence. The gathering of that evidence involves the measurement of frequencies which can be regarded as surrogate approximations of epistemic probabilities, given the Law of Large Numbers. Uncertainty about the future is regarded as a mental state which involves the entertaining of partial degrees of belief about future observations equal to the epistemic probabilities assigned to the possibilities of those observations on the basis of the measured frequencies.

In the wake of quantum mechanics came the concept of *stochastic* physical processes, which are objectively probabilistic. Continuing to employ the metaphysics of possibility, a stochastic analysis of quantum processes with multiple possible outcomes supposes that one of those outcomes will be actualised by virtue of a random selection constrained by the objective probabilities of the possibilities. Those objective probabilities are determined by the Born rule when interpreted as assigning quantum amplitude to the fictional possibilities. As in the case of classical mechanics, stochastic theory interprets uncertainty about the future as the entertainment of partial degrees of belief about alternative possible futures but now the partial degrees of belief are equal to the supposed objective probabilities.

The idea of stochastic processes has widely been accepted as plausible by physicists. It can seem plausible that the half-life of an unstable particle is a mind-independent property of that object. However, an air of mystery surrounds the concept, often referred to as *propensity*. How can propensity be a property of an object? What's the ontic status of propensity?

Hugh Everett III replaced the concept of a stochastic process by that of a *dendritic* process. Consider, for example, Lev Vaidman's *World Splitter* [i]. Connecting to the device with a smartphone, you can choose a setup that will initiate a quantum measurement process with six equal-amplitude outcomes, i.e. a *quantum die*. The concept of a quantum die simply having six outcomes is an idealisation which I'll use for the sake of argument to begin with. Later, I'll consider the implications of abandoning that idealisation.

On "rolling" the quantum die, an Everettian observer fissions into six observers, each seeing one of six different outcomes which are all actual [ii p. 459]. Where is uncertainty to be found? Presumably, Everett thought that it was nowhere to be found, which is why he first entitled his thesis *Wave Mechanics Without Probability*. Presumably, the apparent lack of uncertainty didn't bother Everett; after all, what's uncertainty got to do with *physics*? He was simply suggesting that the histories of quantum processes are typically not linear, they *branch*. They're partially ordered series of events, not well-ordered series. Everett's world wasn't many, it was one; a single branching world which Saunders once appropriately dubbed the *quantum block universe* [iii]. However, I'll continue to use the term *Many Worlds* since it's become virtually ubiquitous and is harmless enough so long as it's qualified in ways which will become clear as we go on.

As you roll Vaidman's quantum die, believing that you'll fission, can you really deny being uncertain about what you'll observe? Many theorists have thought so, including Vaidman himself [iv §3]. In search of pre-measurement uncertainty, others have preferred to replace Everett's concept of fission with those of *overlap* and *divergence*, where the body of an observer at a time is one of a multitude of doppelgängers in erstwhile "parallel" worlds [v][vi][vii]. An observer has a single future but is uncertain as to what they'll observe because uncertain as to which *type* of world they inhabit; whether it's a world where outcome A will occur, or outcome B, and so on. The observer is subject to self-location uncertainty. However, inasmuch as that idea is motivated by trying to fill a lacuna left by supposedly absent pre-measurement uncertainty for fission, it's unnecessary, as we shall see.

Content to do without pre-measurement uncertainty, Vaidman has kept to the traditional path by following Everett in believing that on rolling the quantum die you'll split into six "successors", each in a different branch and each seeing a different outcome. He writes:

> The quantum world splitter lets you enjoy all the possibilities in life with no need to choose. Why choose one, when you can do it all (AT ONCE!) [i] (original emphasis)

The idea is that you decide in advance to act on each of six different enjoyable options according to which number is observed after the measurement. An obvious first objection is to ask in what sense it will be "you" acting in those different futures. Each of the six successors is a different observer seeing a different number and it's logically impossible for them all to be the same observer as you. It's clear that a metaphysics of persistence is needed to make sense of Everettian fission even before considering uncertainty.

Vaidman's term *successors* for post-split observers has generally been used by fission theorists and simply fails to meet this objection concerning personal identity. Note that the problem is avoided for the overlap and divergence interpretations of branching because no splitting occurs. Vaidman asserts that *you* can "do it all at once" but you are not any of your successors.

What's required is known as *stage theory*, which was introduced by Ted Sider in 1996 [viii]. It was first explicitly applied to Many Worlds theory in [ix] and most recently in [x §2.1]. It's often thought that a persisting object is one-and-the-same thing from moment to moment; that's what could be called the folk metaphysics of persistence. However, it's not necessary to think of persistence like that. One can understand the history of an object as consisting of a series of momentary parts: *stages*. What Sider recognised is that an object, at any given moment, can be understood to be a stage of its history and that a persisting object can be understood to be one which has a special relationship with the stages which are called its past and future *temporal counterparts*. A persisting object *was* its past temporal counterparts and *will be* its future temporal counterparts. Contrary to folklore, a persisting object (or observer) doesn't have to be one-and-the-same thing from moment to moment after all. If he were to adopt stage theory, Vaidman could say, without fear of contradiction, that you will be each of six different observers, each seeing a different number after you roll the quantum die.

What's the ontic status of non-present stages on this account? That depends on ones view of the ontic status of past and future states of affairs. On the *eternalist*, "block universe" view, which I suggest is most appropriate for Many Worlds theory, the past and future temporal counterparts of a persisting object will be objects which exist in the past and future of the present object. On non-eternalist views, the present object will bear the temporal relations *was* and *will be* to objects which *did* and *will* exist.

## 2.1 The Logic of Uncertainty

For stochastic theory, a quantum die involves an objectively probabilistic process with six possible outcomes. One of those outcomes will be actualised randomly and each possibility can be assigned an objective, mind-independent probability. To put it another way, the quantum die has a propensity. The propensity is such that each of the six possible outcomes has an equal probability of being actualised. There's long been an aura of mystery about propensity, which I hope to dispel.

For stochastic theory, an observer rolling a quantum die is uncertain about the future for the following reason. As with classical mechanics, uncertainty is understood to be a mental state involving the assignment of subjective probabilities, degrees of belief, to alternative possible futures. Stochastic theorists derive the values for the degrees of belief by appealing to what's become known as the Principal Principle, which is basically the idea that an observer should assign subjective probabilities to possible outcomes equal to what they believe the objective probabilities of those outcomes to be [xi §2.2]. For stochastic theory, the degrees of belief are guided by what are taken to be objective probabilities, whereas for classical mechanics the degrees of belief are guided by epistemic probabilities arising from ignorance of microstates. According to stochastic theory, an observer is uncertain about the future prior to rolling the quantum die because they assign degrees of belief 1/6 to each of the possible outcomes, whose objective probabilities are 1/6.

Vaidman, like other Many Worlds theorists such as David Papineau and Sean Carroll, has followed Everett in understanding the process involved in rolling the quantum die to be dendritic rather than stochastic. All six outcomes occur, each in a different branch of physical reality. Each branch is assigned the same quantum amplitude as is assigned to the *possible* outcomes of stochastic theory and, since the branches actually exist, quantum amplitude must be a physical property which they possess. The absolute square of quantum amplitude is the quantity which stochastic theorists identify with objective probability and that can seem acceptable when amplitudes are assigned to alternative possibilities, but can it be acceptable when amplitudes are assigned to coexistent actualities? Can objective probability be a property of branches?

It's certainly *logically* possible, for if the objective probability of all the outcomes occurring is 1 then that entails that each of the outcomes will occur but it does not give reason to believe that the objective probabilities of each of those individual outcomes must also be 1. The objective probability of the occurrence of each outcome can be 1/6, contrary to the common belief that if an event will occur then the probability of its future occurrence must be 1. That may seem to involve a contradiction because a well-informed observer must be certain that any particular outcome will occur whilst assigning it an objective probability of 1/6. However, the observer is not required to apply the Principal Principle here, where the future *occurrence* of the outcomes is concerned.

There is as yet no agreed justification of the Principal Principle; it's used by stochastic theorists simply because it seems self-evident. If you believe that a process has six possible outcomes whose objective probabilities are 1/6, what else can you do but assign a degree of belief of 1/6 to the future occurrence of any particular outcome? However, stochastic theorists are in the habit of applying this idea in the context of multiple futures thought of as *alternatives*, whereas in the context of the dendritic quantum die the futures are being thought of as coexistent. In that context, the application of the Principal Principle is overruled by logical consequence because, again, if the objective probability of all outcomes occurring is 1 then, necessarily, each outcome will occur, whatever its individual objective probability of occurrence. The observer can assign a subjective probability of 1 to all the outcomes occurring because the objective probability of their combined occurrence is 1. This entails that the observer is certain that each outcome will occur, *despite the objective probability of the occurrence of each outcome being 1/6.*

I should mention in passing that this brings an alternative perspective to the Deutsch-Wallace decision theory argument that observers should assign degrees of belief to future measurement outcomes in accordance with the Born rule [xii pp. 160-89]. If objective probability can be understood to be a property of future branches, then the decision-theoretic argument, if good, constitutes a justification of the Principal Principle and thus belongs to the philosophy of probability rather than to Many Worlds theory.

In what sense, then, can an observer be uncertain about the future prior to rolling the quantum die? They can be uncertain in the sense of assigning a subjective probability of 1/6 to each of the *future observations*. The observer will be each of six observers seeing different outcomes whose objective probabilities are 1/6. Applying the Principal Principle, the observer assigns a degree of belief of 1/6 to the future observation of each outcome. They are certain as to what will *occur* and uncertain as to what they'll *observe*. Whether the future observations are understood to be alternative possibilities or coexistent actualities is beside the point, uncertainty is the very same thing in both cases. The thrall of a folk metaphysics of alternative possibilities can make this hard to grasp.

Should doubt remain, a thought experiment demonstrates that an observer can believe that they are assigning subjective probabilities to alternative possible future observations whilst they are *in fact* assigning them to coexistent actual future observations. This involves a set-theoretic metaphysics for physical objects which leads directly to an explanation of how objective probability can be a physical property of Everettian branches.

## 2.2 Many Worlds without Everett

What cosmologists call the observable universe is a finite region of space which is currently estimated to have a radius of about 46 billion lightyears. Since there is as yet no evidence that space is finite, there may be a denumerably infinite set of such regions which are observationally identical.

Consider an observer who inhabits one of an infinite set of observationally identical universes where quantum dice are, hypothetically, stochastic. On rolling a die, an infinite number of doppelgängers in the set of erstwhile "parallel" universes move in concert and an infinite number of quantum dice are rolled. The set of universes subsequently partitions into six subsets whose measures are *necessarily* 1/6, the reason being that what it *means* in stochastic theory for an outcome of a particular type of process to have an objective probability of 1/6 is that the subset measure for that outcome on an infinite set of such processes is 1/6, the appropriately-named *probability measure*.

Now drop the ubiquitous assumption of folk metaphysics that there's a one-to-one relation between observers and doppelgängers. This requires an exercise in what Donald Davidson has called *radical interpretation* [xiii]. The idea is that truth values must be preserved for relevant utterances by an observer on the original interpretation and the alternative. On the original interpretation of the parallel universes setup, a single utterance by an observer is tokened by a single noise made by a single doppelgänger, but on the alternative interpretation a single utterance is tokened by the infinite number of isomorphic noises emitted by each of the doppelgängers. Likewise for intensional acts: on the original interpretation the act of rolling a die is tokened by the movements of a single doppelgänger, whereas on the alternative interpretation the act of rolling a die is tokened by the parallel movements of all the doppelgängers. On the alternative interpretation, a single die is rolled by a single observer; a single die which is constituted by all the parallel dice. This is the *unitary interpretation of mind* [xiv §2].

A novel use of set theory is required [x §4]. Following Willard Quine, physical objects in each individual observable universe are to be construed as self-membered singleton sets which have come to be known as logicians as *Quine atoms* [xv p. 31]. Quine spent much energy trying to find a way to do mathematical logic without reference to sets, but failed. Having become resigned to the necessity of sets, he noticed that non-sets could be brought into the set-theoretic fold in a way which is harmless in the sense that it doesn't impair the use of set theory in mathematics. Sets had always been understood to be abstract objects but Quine demonstrated that concrete objects could be construed as sets too. What's required for the unitary interpretation of mind is the hypothesis that any set of Quine atoms has all and only the properties its elements share other than number of elements and value-definiteness, when the elements have different definite values. I'll say more about this assumption and its consequences in the next section.

So now there's a single subject whose body is an infinite set of doppelgängers and who rolls a die which is an infinite set of hypothetically stochastic dice. When the single observer rolls the quantum die, each of the doppelgängers which are set-theoretic elements of the observer's body moves isomorphically so that the parallel quantum dice are caused to roll. In each elemental universe the outcome gives rise to sensory input to a doppelgänger so that, as the set of elemental universes partitions into six subsets with different outcomes, so the set of doppelgängers partitions into six subsets with different sensory input. Differences in sensory input give rise to different observations so the single observer fissions into six observers seeing different outcomes. The bodies of the six downstream observers are each an infinite set of doppelgängers whose subset measures relative to the body of the upstream observer are 1/6, i.e. the probability measure.

For this non-Everettian cosmological setup, the single die of the alternative interpretation is not stochastic, it's dendritic. The conclusion must be that an observer can be mistaken when believing that their uncertainty prior to rolling a quantum die derives from there being six *alternative possible* outcomes which all have an objective probability of 1/6. Their uncertainty can derive from there being six *coexistent actual* outcomes which all have an objective probability of 1/6.

## 3. A Metaphysics for Everettian Fission

According to Everett, the quantum die splits into six dice, each showing a different number, and the observer splits along with it. As he saw it, *of course*, there can be no probability since there's no uncertainty, thus his pursuit of a back door to probability via typicality.

Everett's key idea was that the concept of a stochastic process could be replaced by that of a dendritic process. To make it fully intelligible, there has to be an account of how a well-informed observer can be uncertain about future observations in a quantum measurement situation, i.e. observations they will make, together with other nearby observers who have split too, along with the measuring device and the laboratory. We now have an account:

*Uncertainty without alternatives*

> *Uncertainty about future observations is the cognitive state of assigning partial degrees of belief to multiple observations; whether those observations are thought of as alternative possibilities or coexistent actualities is irrelevant because the occurrence of a future observation doesn't entail that the objective probability of its occurrence is 1, so multiple futures can each have probabilities <1.*

If it's useful to our understanding of physics to employ the concept of fission rather than that of stochasticity, then we are free to do so. To be certain that all outcomes will occur entails that each will occur. Therefore we can be certain that any particular outcome will occur whilst believing that the objective probability of its occurrence is 1/6. Assuming the Principal Principle, the observer assigns a degree of belief of 1/6 to the future observation of that outcome, by observers who they and their laboratory colleagues will be.

How can the real-world quantum die split in such a way that the objective probability of each of its immediate future temporal counterparts is 1/6? By being an infinite set which partitions into subsets with probability measures of 1/6. The cosmological thought experiment provides the framework for a metaphysics for quantum fission which incorporates a modification of Quine's definition of concrete objects as being self-membered singleton sets:

*Concrete Sets*

> *A set of Quine atoms has all and only the properties which its elements share with exceptions for number of elements and value-definiteness, when elements have different definite values for some property.*

That this assumption has an Alice in Wonderland aspect, which has long been associated with quantum mechanics, is not to be denied. It entails that the set of a duck and a rabbit, each weighing a kilo, is a *duckrabbit*, a warm-blooded creature which is neither mammal nor fowl and has an indefinite number of feet, though no more than four and no less than two, and which weighs a kilo. To see why, consider isomorphic rooms, one on the left and the other on the right, each containing a doppelgänger. Put the duck and the rabbit in identical boxes and introduce them isomorphically to the rooms, the duck on the left and the rabbit on the right. The single observer's body, the set of the two doppelgänger, makes parallel movements resulting in boxes being moved to scales so that the observer reports that the box which is the set of two boxes contains something which weighs a kilo. It's the duckrabbit. When the observer opens their box they split into Lefty who finds a duck and Righty who finds a rabbit.

If the set of the duck and the rabbit is a third creature which weighs a kilo, is the unit set of that set an object which weighs a kilo too? And the unit set of that set, and so on? There's no good reason to suppose so. Up until now the unit sets of concrete objects, other than Quine atoms, have generally been supposed to be abstract and the Concrete Sets rule doesn't entail that that's not the case; it says nothing about sets of sets of Quine atoms. Set theory is metaphysical Meccano which can be applied to physics in any way we wish, so long as it leads us not into contradiction.

*3.1 From Metaphysics to Physics*

The cosmological thought experiment invokes an infinite set of elemental parallel stochastic universes populated by Quine atoms. However, the whole point of Everett's idea was to *replace* stochasticity with fission. For Everettian physics, the elemental universes must have deterministic, linear histories with branches emerging as the set partitions. Pilot Wave theory provides possible candidate elemental universes [x]. Interacting worlds theory also provides candidate universes with a purely particle ontology [xvi][xvii][xviii], though it may be replaceable by a field ontology [xix]. However, both Pilot Wave and Interacting Worlds theories face problems in dealing with quantum field theory and involve nonlocality in the sense that there can be causal connections between spacelike-separated events.

An often-vaunted advantage of Many Worlds theory is that it doesn't confront those issues. When conceived of, following Everett, as a *pure wave* theory, all of the physics used by physicists can be recovered, so the story goes. In defence of Many Worlds as a pure wave theory, Wallace has recommended a *mathematics-first* approach to the ontology of quantum mechanics, which excludes microscopic local beäbles as objects bearing properties [xx]. The project of ontic structural realism, which he defends, is an interesting one but I suggest that it's better suited to a proto-spacetime ontology than to quantum mechanics, where stuff happens *in* spacetime.

As Louis de Broglie once remarked, a Schrödinger wave is supposedly in configuration space but lacks configurations [xxi p. 381]. There are currently other attempts to fix that by introducing microscopic local beäbles to Many Worlds theory [xxii][xxiii]. What I've been describing is a metaphysical framework which is independent of whatever physics may actually be involved. Assuming a particle ontology, just for the sake of illustration, this framework entails that any macroscopic object in our environment is a set of objects which are macroscopically isomorphic but differ in their microscopic particle configurations. There's a sense in which we inhabit configuration space. Objects in our environment have spatial extension and they're extended in configuration space too, as are our bodies. In effect, the unitary interpretation of mind, is a *consequence* of assuming that objects in our environment are extended in configuration space.

As explained in §2.2, recall that the unitary interpretation of mind is the idea that multiple doppelgängers instance a single observer, not multiple observers in qualitatively identical mental states. If your body is understood to be extended in configuration space, in the sense of being a set of bodies that are only anisomorphic at the level of microscopic configurations, then your mental state, now, is instanced by a multiplicity of doppelgängers. You are legion, to adapt a biblical phrase.

In light of this, think about Vaidman's quantum die again. It's an apparatus in a physics lab which is a set of labs including all possible configurations of particles consistent with the Born rule. That's the reason why the set which is the die partitions in the same way that a set of hypothetically stochastic dice would. However, is Vaidman's die an infinite set? That would depend on whether spacetime is continuous. Can the branch subset measures still be identified with objective probabilities if the set of deterministic quantum dice is finite? Perhaps not, in which case perhaps an effective Law of Large Numbers is good enough for very large samples. In any case, given the cosmological setup, if there's a finite number of configurations there can still be a denumerably infinite set of each configuration until such time as we have evidence that space is finite, because there can be an infinite number of observable universes which are observationally identical.

According to this framework, an unstable particle in our environment would be a set of particles constantly partitioning into a decay subset of increasing measure. An observer with a detector would be constantly splitting into an observer not seeing decay and observers seeing decays at later and later times. The probability of observing decay within a given period would depend on the rate of change of the decay subset measure for that type of particle, i.e. its propensity to decay. Likewise, we are free to hypothesise that the quantum die is a very large or infinite of set of dice which will partition in the same way as a set if stochastic dice would. The subset measures of Vaidman's six downstream dice will be 1/6 relative to the upstream die, because that's the probability measure.

For another illustration of the idea that objects in our environment are extended in configuration space, consider a free electron at any given moment. It's a set of elemental electrons which are in different corresponding positions and have different momenta in the elemental universes. The term *elemental* here is strictly set-theoretic. Again, our universe is being construed as a set of universes and any object in an observer's environment is a set of Quine atoms. A free electron in our universe is a set of elemental electrons which are on different trajectories in the universes which are elements of our universe. Our electron doesn't lack a trajectory, it has an *indefinite* trajectory. At any moment, the electron has indefinite position and momentum in our *non-elemental* universe, where objects have a definite position and momentum only if their elements have corresponding positions and momenta in the elemental universes. There's more on this concept of correspondence in §4.1.

The introduction of particles as local beäbles in the way I've described, as being set-theoretic elements of particles in our environment, effectively preserves the full structure of the wavefunction and avoids the drawbacks of Pilot Wave and Interacting Worlds theories, as I shall now explain.

## 3.2 The World as Wavefunction

Consider the wavefunction of a free electron understood in terms of the set-theoretic metaphysics. For a pure wave theory, an observer assigns any region of space a quantum amplitude and the absolute square is taken to yield the *probability* of finding the electron there if a position measurement is made. There's no account of how an *actual* electron can be "spread out" in this way, hence Wallace's appeal to a thingless ontology. However, for the set-theoretic metaphysics the absolute square of amplitude for a spatial region yields a subset measure for the single free electron, which is a set of elemental electrons. Each elemental electron in that subset is at an elemental location which is an element of a location within the given spatial region. There is thus a *fully concrete* interpretation of the wavefunction in that region for the free electron in an observer's environment, which I shall call an *environmental* electron. It's not in any sense counterfactual. Every location in that region is a set of elemental locations where elemental electrons may be *actually located*. But there's no sense in which a *part* of the environmental electron is located in an observer's spatial region. The *environmental* electron has no actual presence in an observer's spatial region; it's rather that the *elemental* electrons which are set-theoretic elements of the *environmental* electron are actually present at *elemental* locations which are elements of an *environmental* location. I shall continue to italicise the terms *environmental* and *elemental* for the sake of clarity in what follows.

Note that this appears to invoke a substantivalist interpretation of spacetime. To construe the universal wavefunction as being a set of universes "extended" in configuration space entails that an *environmental* spacetime region is a set of *elemental* spacetime regions. An *environmental* event is a set of corresponding *elemental* events, each in an *elemental* universe, one where there's a particular configuration of microscopic local beåbles on any given simultaneity hyperplane. That implies that Everettian fission involves the fission of spacetime itself. I'll say more about this in the following section.

It's often said that the paradox of superposition is dealt with in Many Worlds theory by understanding superpositional states as being composed of definite states on different branches. Thus Schrödinger's cat is dead on some branches and alive on others (sometimes put as dead in one world and alive in another). However, Everettian theory has only ever given an account of *macroscopic* superpositions in this way. Mystery still surrounds the concept of microscopic superpositons, hence, again, the motivation for defending a pure wave theory in terms of an ontology that doesn't involve objects bearing properties. The set-theoretic metaphysics resolves this problem by construing microscopic superpositions as also being constituted by multiple definite states. Again, the free electron in an observer's *environment* becomes an extended object, extended in configuration space. It does so by being a set of electrons, each of which is on a different trajectory in a universe which is a set-theoretic element of the observer's universe.

But that only provides a momentary snapshot of the electron's wavefunction. There needs to be the dynamics of unitary evolution too; where is that to come from? It strikes me that the most plausible option here is to adopt some version of Interacting Worlds theory. The individual elemental universes which contain the set-theoretic elements of the observer's electron interact in such a way as to generate the unitary dynamics. Here, there's scope for new physics in order to understand how universes which are set-theoretically separated in configuration space interact. The possibility of such new physics has already been suggested by Interacting Worlds theorists, but what must be stressed is the radically new perspective that the set-theoretic metaphysics brings to Interacting Worlds theory.

All the difference is in how the observer is situated. For extant Interacting Worlds theory the observer is situated within an individual world, which corresponds to what I've been calling an *elemental* universe. For the set-theoretic metaphysics, the observer is situated in the *set* of interacting universes; objects in the observer's environment, including their body, are multipleton sets. The observer's universe becomes a set of interacting universes.

In a sense, the observer spans the set of interacting universes. They span the universes in the sense that the mental states of an observer are instanced by a multitude of brains in a multitude of doppelgängers. Each of those brains is a set-theoretic element of the brain to which an observer indexically refers by a tap on the skull. The observer's mental states are instanced by a multitude of brains rather in the way that a single novel is instanced by a multitude of books.

Extant Interacting Worlds theory involves causal nonlocality because particle trajectories in the observer's world are mutually interactive at spacelike separation in virtue of the interactions between worlds. By construing our universe as a set of interacting universes rather than an element of a set, nonlocality is avoided. Causal locality is preserved for Many Worlds theory, as we shall see.

## 4. Being Indefinite

Consider an observer who rolls a quantum die blindfolded. According to Vaidman, the observer will fission into six successors, each on an Everettian branch where the outcomes are different. According to the set-theoretic metaphysics, the body of the observer will partition into six subsets and each subset will have elements which are doppelgängers in the presence of elements of one of the six outcomes. The partitioning of the observer's body will be caused by slight physical effects propagating from the six different post-roll dice, even if those effects are very slight indeed, such as gravitational differences. However, the observer themself won't fission because the doppelgängers aren't different enough to instance distinct perceptual states. The observer doesn't fission because their perceptual apparatus is screened by the blindfold. Post-measurement and pre-observation there'll be a single successor whose body is the set of all the doppelgängers in the six subsets. The environment of that single successor will contain a die with subsets which are six dice displaying different numbers. In other words, the die in the environment of the post-measurement, pre-observation observer will be in an *indefinite* number-state [xiv p. 14].

Now consider a terrestrial observer watching the roll of a quantum die on Mars through a powerful telescope. Post-roll on Mars, there'll be no causal influence on the observer's body on Earth for several minutes and so there'll be no relevant partitioning of the observer's body. When light from the roll of the die reaches the observer's eyes, their body will partition into six subsets and then, after retinal states have been processed, there will be six sets of doppelgängers instancing elements of six different perceptual states and the observer will have fissioned. During the intervening few minutes, the quantum die will have been in an indefinite number-state relative to the terrestrial observer but not relative to a Martian observer.

For the set-theoretic metaphysics, an observer cannot fission into observers seeing different outcomes until the observer's body partitions into subsets which are bodies instancing different cognitive states. Note that this has nothing to do with *consciousness*, it has to do with *mental content*. It's well established by experimental psychology that we can perceive states of the world around us whilst not being conscious of those perceptions; and we can recall information to consciousness which we were not conscious of knowing a moment earlier. Two distinct observers may be in qualitatively identical conscious states and yet act differently because of different unconscious mental content.

Necessarily, quantum measurements with multiple outcomes which occur at spacelike separation from an observer are in indefinite outcome states relative to that observer. This casts doubt on the idea that the observation of correlations between spacelike-separated measurements on entangled particles entails nonlocal causation. That conclusion only necessarily follows if measurements always have single, definite outcomes. However, the set-theoretic metaphysics construes the observer's universe as a set of universes and within the elemental universes there seems to be nonlocality for the reasons acknowledged by Pilot Wave and Interacting Worlds theorists. So, is nonlocality involved on the set-theoretic view after all?

No, because apparent nonlocality at the *elemental* level is not really nonlocality at all. It would be if observers inhabited the individual *elemental* universes but the whole idea is that they do not. Observers inhabit sets of *elemental* universes and, at that level, nonlocality would seem to be absent. *Elemental* nonlocality is not nonlocality because *elemental* locations are not locations. For the set-theoretic metaphysics, there's no reason to suppose that there's causal influence between spacelike-separated *locations*, which are locations in an observer's spacetime, a set of *elemental* spacetimes. This will become clearer with an analysis of EPR-Bell experiments, and what's needed by way of preparation for that is the set-theoretic characterisation of spin and entanglement coming in the following subsections, but first a few words on what may seem a contradiction.

The Concrete Sets rule is that sets of concrete objects have all and only the properties which their elements share other than number of elements and value-definiteness. How then can there be no nonlocality in a set of universes whilst there seems to be a form of nonlocality in the universes which are its elements? The reason is that the locality issue is tied to the concept of value-definiteness, as we've just seen, and of which there's more to come.[1]

*4.1 Spin*

Spin poses a further challenge to the set-theoretic metaphysics. We have to take a step back. The *environmental* universe is being construed as a set of *elemental* universes. An *environmental* electron only has a location if all the *elemental* electrons which are its elements are at corresponding *elemental* locations.

The correspondence can be thought of in the following way. For an observer at a time, the *environmental* universe exhibits a definite distribution of objects in space on the surface of the past light cone. The observer's universe at a time is to be construed as the set of universes containing all possible configurations of particles (or fields) consistent with that definite distribution of objects. An *environmental* particle only has a position in the observer's universe if its elements are all at the same position relative to isomorphic distributions of macroscopic objects in each *elemental* universe.

The set-theoretic metaphysics interprets objects with indefinite properties as sets of objects with definite properties, so, when it comes to spin, *elemental* electrons cannot have indefinite spin relative to any orientation. An *elemental* electron must have a definite spin, i.e. *up* or *down*, relative to some orientation, period. Just as an *environmentntal* free electron has an indefinite trajectory whilst the electrons which are its elements follow trajectories, likewise, an *environmental* electron can have indefinite spin relative to all orientations but one whilst the electrons which are its elements simply have definite spin relative to a single orientation. Observers always make measurements on *environmental* electrons which are sets of *elemental* electrons (which are Quine atoms, like all physical objects in an *elemental* universe). The spin of an *elemental* electron can't be measured. Quantum measurement is something we do in our *environmental* universe but not in the *elemental* universes which are its set-theoretic elements. Bearing that in mind, here's an attempt to provide a set-theoretic metaphysics for spin.

---

[1] My thanks to Rainer Plaga for raising this point.

Following the Concrete Sets rule, a z-spin-up *environmental* electron must be a set of *elemental* electrons for which all the z-spin *elemental* electrons are z-spin-up. For any orientation ô, other than the z-axis, there will be a subset of elements of the *environmental* z-spin-up electron which are *elemental* electrons with definite spin relative to ô. Furthermore, for that subset, the measures of the ô-spin-up and ô-spin-down *elemental* electrons must be the probabilities for measuring ô-spin-up and ô-spin-down if the ô-spin of the *environmental* z-spin-up electron is measured. For the special case of the x-axis the *environmental* z-spin-up electron will have a subset of *elemental* x-spin electrons which has equal measures for its x-spin-up and x-spin-down elements. Likewise for the y-axis. Thus the body of an observer measuring the x-spin or y-spin of the *environmental* z-spin-up electron will fission into bodies of equal measure of observers seeing opposite x or y spins.

Why do the post-measurement observers have bodies of equal measure? Recall the cosmological thought experiment with an infinite set of hypothetically stochastic universes. Now think of an equal-chance measurement being made in each universe, i.e. a quantum coin flip. The set of universes will partition into two subsets of equal probability measure where different outcomes occur. For that setup, if the unitary interpretation of mind is adopted, there's a single observer at the outset whose body is a set of bodies (doppelgängers) that partitions into two subsets of equal measure, which are the bodies of the two post-coin-flip observers. Recall also that in §3.1 the set of hypothetical stochastic universes was replaced by a set of Pilot Wave or Interacting Worlds universes, which would partition in the same way as a set of stochastic universes would, i.e. the branch subset measures would take the same values. The reason for this would be that the set of *elemental* universes would include all possible *elemental* configurations consistent with the Born rule (corresponding to the assumption of "equilibrium" in Pilot Wave theory). To put it another way, the universal wavefunction is being interpreted as a set of universes which includes all possible configurations and so Everettian branching is construed as the partitioning of a set, where the subset measures just *are* the outcome probabilities. To repeat, because it's very counterintuitive, the perspective is that which arises from the unitary interpretation of mind coupled with the idea that *environmental* objects are set-theoretically extended in configuration space by being sets of *elemental* objects, so the fissioning of an observer consists in the partitioning of the observer's body into subsets.

For the intermediate ô-spin cases the measures of the ô-spin-up and ô-spin-down outcome branches will be equal to the probabilities entailed by quantum mechanics because the *environmental* z-spin-up electron is *constituted* by a set of *elemental* electrons such that the subset of ô-spin *elemental* electrons has the appropriate measures for ô-spin-up and ô-spin-down. This provides a characterisation of spin for the set-theoretic metaphysics, however, before we can apply it to the analysis of EPR-Bell experiments we need a set-theoretic characterisation of entanglement.

*4.2 Entanglement*

A pair of electrons in a singlet state has zero net spin because they have opposite spins. Emitted from a source and collimated, the wavefunction propagates as a sphere with peaked amplitudes in opposite directions. The wave propagates in configuration space but the set-theoretic metaphysics provides, at any given moment, a characterisation of the wave as a distribution in 3D space, but not of anything which *exists* in 3D space. Rather, it's a distribution of *elemental* objects, each in an *elemental* universe, which have *elemental* positions that are set-theoretic elements of the 3D *environmental* positions.

Both the *environmental* entangled electrons are sets of *elemental* electrons. At any given region of *environmental* space at a time (the space in the environment of the observer), there will be subsets of the *elemental* electrons which are elements of the *environmental* electrons and which are located in *elemental* regions which are the set-theoretic elements of the given *environmental* region. This is what I meant when I said earlier that the set-theoretic metaphysics completely recovers the structure of the wavefunction. Here, we see an instantaneous reconstruction (relative to some simultaneity hyperplane). The dynamics, which provides the phase aspect of the wave, might be recovered via an Interacting Worlds theory, as I've already suggested.

Consider congruent *environmental* spatial regions of measurement, A and B, which are equidistant from the source and spacelike-separated. Both *environmental* regions are sets of *elemental* regions containing electrons which are elements of each of the two entangled *environmental* electrons. Since electrons lack haecceity, there's no sense in which the entangled electrons can be permutated, but there are two of them nonetheless, so there are two distinct sets of *elemental* electrons which are the elements of the entangled *environmental* electrons.

Furthermore, for each of the two *environmental* regions there will be *elemental* regions containing elements of each of the entangled *environmental* electrons such that for every orientation there's a subset of an *environmental* electron with definite spins for that orientation and with equal measures for spin-up and spin-down, because if either *environmental* electron has its spin measured relative to any axis there are equal probabilities for spin-up and spin-down. Both the entangled *environmental* electrons will be equally present in both *environmental* regions, A and B, so to speak, where the presence of a free electron in an *environmental* region is construed as its having subsets of *elemental* electrons which are located in *elemental* regions which are elements of that *environmental* region. So the two entangled *environmental* electrons are *separable* because they're two distinct objects. They are two distinct sets of *elemental* electrons with no elements in common. There's a single wavefunction for two distinct *environmental* electrons. This analysis clarifies the account I gave in [$^{x}$ §3] which wrongly assumed that the entangled electrons are non-separable.

*4.3 EPR-Bell*

We are to consider Alice and Bob making spin measurements on the singlet state in regions A and B. When Alice makes her measurement on orientation ô she fissions into Alice$_{UP}$ and Alice$_{DOWN}$ whose bodies occupy the local *environmental* spatial regions A$_{UP}$ and A$_{DOWN}$, which are subsets of region A. The set of the *elemental* points which are the elements of points in A is the fusion of the two distinct subsets whose elements are *elemental* points in A$_{UP}$ and A$_{DOWN}$. The fissioning of Alice's body involves the fissioning of the *environmental* spatial region which it occupies. Prior to measurement, Alice inhabited an *environmental* region which was a set of *elemental* regions, each in an *elemental* universe. Post-measurement, Alice$_{UP}$ and A$_{DOWN}$ inhabit two distinct *environmental* regions which contain *elemental* points in two distinct subsets of *elemental* universes.

What distinguishes the *environmental* regions A$_{UP}$ and A$_{DOWN}$ is that they contain two different *environmental* electrons and two different subsets of the *elemental* bodies which are the set-theoretic elements of the bodies of Alice$_{UP}$ and Alice$_{DOWN}$. Region A$_{UP}$ contains all the elements of Alice$_{UP}$'s body and none of the elements of Alice$_{DOWN}$'s, and vice versa. In Bob's absolute elsewhere, Alice's body has evolved into an indefinite ô-spin-measurement state with equal measures for ô-spin-up and ô-spin-down because it has partitioned into subsets which are the bodies of Alice$_{UP}$ and Alice$_{DOWN}$, in different branches of the quantum multiverse.

Keeping things simple to begin with, let Bob make his measurement on orientation ô-too. He fissions into Bob$_{UP}$ and Bob$_{DOWN}$ in regions B$_{UP}$ and B$_{DOWN}$. The key point here is that, because of the entanglement, these two subsets of region B have different set-theoretic structures from the regions A$_{UP}$ and A$_{DOWN}$ whilst regions A and B are set-theoretically isomorphic. *Necessarily*, Alice$_{UP}$ cannot have measured the *same* electron as Bob$_{UP}$ and Alice$_{DOWN}$ cannot have measured the same electron as Bob$_{DOWN}$. That's a consequence of the two entangled *environmental* electrons having opposite spins relative to any orientation.

Now Bob's body has evolved into an indefinite ô-spin-measurement state relative to both Alice$_{UP}$ and Alice$_{DOWN}$ and Alice's body has evolved into an indefinite ô-spin-measurement state relative to both Bob$_{UP}$ and Bob$_{DOWN}$. These measurement results of the Alices and Bobs can't come into causal contact sooner than half the light-time between regions A and B. To see why, consider Clotilde, halfway along a light path between regions A and B and watching the observers who will be the Alice and Bob who make the measurements. When Clotilde sees the results of Alice's and Bob's measurements she fissions into Clotilde$_{AliceUP+BobDOWN}$ and Clotilde$_{AliceDOWN+BobUP}$. As Cai Waegell and Kelvin McQueen put it, "A world containing a Bob and an Alice is only created when the wavefront from Alice's measurement meets the wavefront from Bob's measurement" [[xxiv]§6]. However, it's unclear why they use the term *wavefront*; it's rather a matter of the past light cones of Alice's and Bob's future temporal counterparts coming to overlap.

Things get more complicated if Bob makes his spin measurement for a different orientation from Alice and it's here that an *apparent* non-locality shows up, but it's "nonlocality" at the *elemental* level, not nonlocality at the *environmental* level. It's for this reason that experiments involving the changing of spin measurement orientations for Bob are important, because the correlations reveal a more detailed picture of what's going on.

When Bob makes his measurement his body partitions, *but it doesn't partition into subsets of equal measure*. The different measures correspond to the different probabilities for measuring spin-up and spin-down for the chosen orientation, *given Alice's measurement*. So the set-theoretic structure of region B must be caused to change as a result of Alice's measurement in region A. On reflection, that should be expected. As I mentioned earlier, Pilot-Wave universes are appropriate candidates for being *elemental* universes and for Pilot-Wave theory entangled particles are mutually influential at spacelike separation.

However, a change in the set-theoretic structure of an *environmental* region isn't a change in anything which exists *in* that region. *Elemental* objects do *not* exist in that region, they exist in *elements* of that region. The set-theoretic metaphysics reveals how the EPR-Bell correlations arise without anything which exists in the environment of an observer being caused to change at spacelike separation. *Something* changes in region B when Alice makes her measurement in region A, but not anything which *exists* in region B, only things which exist in the regions which are set-theoretic *elements* of region B. The set-theoretic metaphysics furnishes separability and thereby provides a mechanism which explains the EPR-Bell correlations without invoking nonlocality. That mechanism operates at the *elemental* level.

## 5. Beyond Idealisation

With the set-theoretic metaphysics in place, consider a non-idealised version of Vaidman's quantum die. Apart from the six equiprobable outcomes, there will be a plethora of extremely low-amplitude outcomes. Outcomes where "quantum accidents" occur, such as your smartphone transforming into a simulacrum of a salamander rather than displaying one of six numbers. These sorts of future events were also conceivable for classical physics as the result of highly improbably particle trajectories, amusingly illustrated in Bertrand Russell's tale *The Metaphysician's Nightmare* [xxv]. However, on the fission interpretation of Many Worlds theory, all such bizarre events exist in the multiple futures of an observer. Vaidman doesn't take them into account because such events have, as he would put it, very low *measure of existence* [xxvi]. I've effectively argued that Vaidman's "measure of existence" can be strictly identified with objective probability. So bizarre futures should be left out of account when rolling a quantum die because they have ridiculously low probabilities. There's nothing new in that idea.

However, the idea that all those bizarre futures *actually exist* is not necessarily anodyne. Pause for thought is called for in view of scenarios such as Huw Price's *Legless at Bondi* [xxvii p. 382]. More briefly, suppose that you're ill and offered treatment which involves quantum processes with multiple outcomes. There's a high probability that you'll be cured but a low probability that you'll end up much worse off. In a conventional context you may well take the risk, even if a little anxiously. In the fission context, you can be sure that the cured person will know that someone else is suffering because of the decision you took. Is it consolation enough to know that the suffering person will also have been the person who took the decision? It's not obvious that a fission interpretation of Many Worlds is free from moral conundrums. Another interesting case is to be found in [xxviii]. But then why should we expect such a profound change of worldview not to have consequences for how we understand the human predicament?

**6. Parting Lines**

I've argued that Everett's key idea was to replace the concept of a stochastic process with that of a dendritic process, which is the thought that quantum phenomena induce the splitting of observers and their environments. This ostensibly raises problems which cannot be resolved by physics alone because assumptions rooted in folk metaphysics stand in the way. Observers cannot make predictions and test them unless they persist, but how can an observer persist through fissioning into multiple different observers? Sider's stage theory solves that problem, but it didn't become available until 1996 and remains neglected in the philosophical literature on persistence.

How can an observer be uncertain about future observations whilst believing that multiple outcomes will occur? The folk metaphysics of possibility and actuality stands in the way but logic doesn't. That the objective probability of multiple outcomes occurring is 1 does *not* entail that the objective probability of each outcome's occurrence is 1. In that case, uncertainty can be understood as assigning partial degrees of belief to multiple future observations without those future observations needing to be alternative possibilities, as has always been thought. They can be coexistent actualities.

How can objective probability be a property of multiple actual outcomes? The proposal which has been described and further developed here involves the hypothesis that individual objects in an observer's environment can be construed as multipleton sets which are macroscopically isomorphic and microscopically anisomorphic because they're constituted by different configurations of local beäbles. Quantum processes induce the partitioning of those sets into macroscopically distinct subsets whose measures are the objective probabilities of outcomes. As a consequence, a single observer's body is a set of doppelgängers, so the belief that there can be multiple copies of *observers*, which is widely held, should be rejected. A future-oriented account of objective probability is provided by the idea that a single observer, whose body is a set of doppelgängers, fissions into multiple observers whose bodies are subsets of doppelgängers with probability measures. According to stage theory, the pre-measurement observer bears the relation *will be* to each of the post-measurement observers and is uncertain about what will be observed because assigning degrees of belief to future observations equal to the probability measures of the future branches. There's no question as to *which* post-measurement observer the pre-measurement observer will be, they will be *each* of them.

This set-theoretic metaphysics provides a framework for a version of the Many Worlds interpretation of quantum mechanics which includes causal locality, separability and Everettian fission (rather than divergence). It provides an account of probability that doesn't appeal to self-location uncertainty and an account of microscopic reality which includes local beäbles. It leaves work to be done on the physics of those beäbles and the way that they participate in the unitary dynamics.